\documentclass[fleqn,usenatbib]{mnras}

\usepackage[T1]{fontenc}

\DeclareRobustCommand{\VAN}[3]{#2}
\let\VANthebibliography\thebibliography
\def\thebibliography{\DeclareRobustCommand{\VAN}[3]{##3}\VANthebibliography}

\usepackage{dcolumn}% Align table columns on decimal point
\usepackage{bm}% bold math
\usepackage{hyperref}% add hypertext capabilities
\usepackage[mathlines]{lineno}% Enable numbering of text and display math

\usepackage{graphicx}	% Including figure files
\usepackage{amsmath}	% Advanced maths commands
\usepackage{amssymb}	% Extra maths symbols
\usepackage{newtxtext,newtxmath}
\DeclareMathOperator*{\argmax}{argmax}
\DeclareMathAlphabet{\mymathbb}{U}{BOONDOX-ds}{m}{n}
\title[Generative modeling of Galactic dust]{A Generative Model of Galactic Dust Emission Using Variational Inference}

\author[B. Thorne et al.]{
Ben Thorne,$^{1}$\thanks{E-mail: blthorne@ucdavis.edu}
Lloyd Knox,$^{1}$
and Karthik Prabhu$^{1}$
\\
$^{1}$Department of Physics, University of California, One Shields Avenue, Davis, CA 95616, USA\\
}

\date{Accepted XXX. Received YYY; in original form ZZZ}

\pubyear{2021}

\begin{document}
\label{firstpage}
\pagerange{\pageref{firstpage}--\pageref{lastpage}}
\maketitle

% Abstract of the paper
\begin{abstract}
Emission from the interstellar medium can be a significant contaminant of measurements of the intensity and polarization of the cosmic microwave background (CMB). For planning CMB observations, and for optimizing foreground-cleaning algorithms, a description of the statistical properties of such emission can be helpful. Here we examine a machine learning approach to inferring the statistical properties of dust from either observational data or physics-based simulations. In particular, we apply a type of neural network called a Variational Auto Encoder (VAE) to maps of the intensity of emission from interstellar dust as inferred from {\it Planck} sky maps and demonstrate its ability to a) simulate new samples with similar summary statistics as the training set, b) provide fits to emission maps withheld from the training set, and c) produce constrained realizations. We find VAEs are easier to train than another popular architecture: that of Generative Adversarial Networks (GANs), and are better-suited for use in Bayesian inference. 
\end{abstract}

% Select between one and six entries from the list of approved keywords.
% Don't make up new ones.

%%%%%%%%%%%%%%%%%%%%%%%%%%%%%%%%%%%%%%%%%%%%%%%%%%

%%%%%%%%%%%%%%%%% BODY OF PAPER %%%%%%%%%%%%%%%%%%

\section{Introduction}
\label{sec:introduction} 

Among the many research enterprises stimulated by the detection of large-scale anisotropies in the cosmic microwave background (CMB) by the COsmic Background Explorer (COBE) with its Differential Microwave Radiometer \citep{smoot92}, is the hunt for signatures of primordial gravitational waves (PGW). To date, only upper limits have been set, most commonly expressed as limits on the ratio of primordial tensor perturbation power to scalar perturbation power, $r$. Soon after the COBE detection it was realized that reliably detecting levels below $r \simeq 0.1$ could not be done with temperature anisotropies alone \citep{knox94}, and that proceeding further would require highly sensitive measurements of the polarization of the CMB on angular scales of about a degree, or larger \citep{kamionkowski97,seljak97}. 

    Polarized emission from the interstellar medium of the Milky Way, in the cleanest parts of the sky at the cleanest observing frequencies, is comparable to the cosmic microwave background signal generated by PGWs if the PGW signal is near the current 95\% confidence upper limit of $r < 0.06$ \citep{bkp/constraints:2018}. So-called Stage III CMB experiments, such as the Simons Observatory \citep{sogoals:2019}, and BICEP Array \citep{hui/etal:2018} combined with SPT-3G \citep{benson/etal:2014} are designed to have sufficient sensitivity and systematic error control to tighten the 95\% confidence upper limits by a factor of about 20. The Stage IV experiments LiteBIRD and CMB-S4 are targeting upper limits factors of 2 and 5 times more stringent still, respectively. Thus we are rapidly moving into a regime where the foreground contamination is up to two orders of magnitude larger\footnote{This is for fluctuation power. The rms level of contamination in the map is up to one order of magnitude larger than the signal of interest.} than the signal of interest. 

The most exciting possibility is that there will be a detection of PGW, as opposed to improved upper limits. A detection claim would essentially be a claim that there is power remaining in the map that cannot be explained as a residual instrumental systematic or residual foreground emission. Detection, therefore, requires not only foreground cleaning, but the capability to quantify the probability distribution of residual foreground power. Such capability is hampered by our lack of prior knowledge of the probability distribution of the non-Gaussian and non-isotropic galactic foreground emission. 

The state of the art in analysis of such observations either implicitly or explicitly has the galactic emission, or their residuals, modeled as Gaussian isotropic fields \citep{planck/parameters:2018,aiola/etal:2020,bkp/constraints:2018}. They are modeled as such not because they are, but strictly for convenience. 

At the very least, we need sufficient simulations of galactic emission to test such algorithms for bias. 
A more ambitious objective is to abandon assumptions of Gaussianity and isotropy altogether, and perform a complete Bayesian analysis with incorporation of an appropriate prior for the spatial distribution of interstellar emission. Groundbreaking progress toward such a Bayesian analysis has been made recently, with the development of analysis methodologies by \cite{millea20a}, and the recent application to real data \citep{millea20b}. 

The analysis framework in \cite{millea20a} was developed for ``de-lensing'' of the CMB; i.e., taking into account the impact of gravitational lensing on the statistical properties of CMB polarization. Although it has not been applied to multi-frequency data, or used for foreground cleaning, at a conceptual level the framework can be straightforwardly extended to analysis of foreground-contaminated multi-frequency data. Although this extension could be implemented with isotropic Gaussian priors for foreground emission, it also presents the opportunity to incorporate more realistic priors -- priors that more accurately reflect what we know about such emission from other data, or from physics-based simulations. 

We are thus interested in both creating simulated maps of galactic emission with the appropriate statistical properties for testing analysis algorithms to be used on real data, and also in learning, from other data and perhaps physical modeling (e.g. MHD simulations of the interstellar medium \citep{kim/etal:2019}) the statistical properties of maps of galactic emission for use in Bayesian inference engines.

Here we report on progress toward accomplishing both of these tasks with the use of neural networks. 
 \cite{aylor/etal:2019} studied the use of generative adversarial networks (GANs) for learning how to simulate new emission maps with statistic properties similar to those from a training set, whilst \cite{krachmalnicoff/puglisi:2020} trained to simulate non-Gaussian small-scale polarized dust emission. Here we present a similar study, this time using a different neural network architecture and training program, that of variational auto encoders (VAEs). 

VAEs and GANs are examples of deep generative models. These models have had recent success in accurately modeling complicated, high-dimensional, datasets, and generating realistic novel samples \citep{razavi/etal:2019,vandenoord/etal:2016,brock/etal:2018}. Generative models can be divided into two main categories: likelihood-based models that seek to optimize the log likelihood of the data, these include the VAE \citep{kingma/welling:2013,rezende/etal:2014}, \emph{flow} based methods \citep{dinh/etal:2014, dinh/etal:2016, rezende/etal:2015,kingma/dhariwal:2018}, and \emph{autoregressive} models \citep{vandenoord/etal:2016b}; and implicit models, such as GANs \citep{goodfellow/etal:2014}, which train a generator and discriminator in an adversarial game scenario. There are many trade-offs to consider when selecting a likelihood-based approach \citep{kingma/dhariwal:2018}, but here we choose to explore the use of VAEs due to their simplicity and computational scalability to higher resolution datasets.

We find some advantages of VAEs over GANs.
The adversarial training process does not produce an explicit inference model, and it is hard to consistently compare model performance against some test set. Furthermore, it is also a common problem that samples from GANs do not represent the full diversity of the underlying distribution \citep{grover/etal:2018}. In contrast, VAEs optimize the log likelihood of the data. This means both that it is possible to directly compare models, and trained models should support the entire dataset, which is crucial when applying a trained model to real data. VAEs also tend to be easier to train in that training success is more stable to variation of hyperparameters. As a downside, VAEs are well known for loss of resolution. We see this in our results and discuss adaptations one could make to avoid this degradation of angular resolution. 

Although our work is motivated by the PGW-driven desire to understand the statistical properties of polarized foreground emission, in this paper, as was the case in \cite{aylor/etal:2019}, we restrict ourselves to intensity. Observations of polarized dust emission with high signal-to-noise over a large fraction of sky do not currently exist, which precludes the training of similar models on real data. However, in ongoing work, we are exploring the use of magnetohydrodynamical (MHD) \citep{kim/etal:2019} simulations to train generative models of polarized emission. In this scenario a trained model would provide a `compression' of the information available in MHD simulations into a single statistical model, which could then be used either in inference, or to augment real low-resolution observations with physically-motivated small-scale realizations. 

The rest of this paper is structured as follows.  In Section~\ref{sec:variational_autoencoders} we introduce variational autoencoders, and the objective for their optimization. We then describe the network architecture we used, the training dataset we produced to train the network, and how hyperparameter values were set. In Section~\ref{sec:results} we present the results of applying the trained VAE to test set images. Finally, in Section~\ref{sec:conclusions} we summarize our findings and discuss areas of current and future work.

\section{Variational Autoencoders}
\label{sec:variational_autoencoders} 

In this Section we will introduce the idea of variational autoencoders, the specific model we implement, and the details of how we train that model.

Our goal here is to take a set of images of thermal emission from interstellar dust $\mathbf{x}^{(i)} = (x_1^{(i)}, \dots, x_N^{(i)}) \in \mathbb{R}^N$, and infer from them an underlying distribution, $p(x)$ from which they could have been drawn, using the techniques of \emph{generative modeling}. Variational autoencoders are a type of generative machine learning model, which provide a framework by which we may infer the parameters of a joint distribution over our original data, and some \emph{latent variables}, $\mathbf z$, representing the unobserved part of the model. We can factorize the joint distribution of the data and latent variables into two terms representing the generative process of the data, and the latent space, responsible for the variance in the observed data:

\begin{equation}
    p(\mathbf x,\mathbf z) = \underbrace{p(\mathbf x|\mathbf z)}_{{\rm Generative}}\underbrace{p(z)}_{{\rm Variance}}.
\end{equation}

The VAE approach is to model the conditional distribution with an appropriate family of functions with some unknown weights, $\theta$: $p_\theta(\mathbf x | \mathbf z) \approx p(\mathbf x | \mathbf z)$. This conditional model encodes the generative process by which  $\mathbf{x}$ depends on the latent set of variables $\mathbf{z}$. The choice of $p(\mathbf z)$ can then be a simple, perhaps Gaussian, prior probability distribution $p(\mathbf z)$, which encodes the dataset variation in a simple latent space. This can be seen as a type of regularization by which we separate out different sources of variation within the dataset, a process that is quite natural for physical processes, and often makes the resulting model interpretable. 

The goal of training is thus to find a transformation that delivers an acceptable approximation $p_\theta(\mathbf x) \approx p(\mathbf x)$,  that is optimal (in some sense), given the training set data. Toward that end we consider the parametrized joint distribution of $\mathbf x$ and $\mathbf z$: 
\begin{equation}
    p_\theta(\mathbf x,\mathbf z) = p_\theta(\mathbf x|\mathbf z)p(\mathbf z),
\end{equation}
which leads to our object of interest via marginalization over $z$:
\begin{equation}
\label{eq:likelihood}
    p_\theta(\mathbf x) = \int d\mathbf z~ p_\theta(\mathbf x, \mathbf z).
\end{equation}

Our tasks are thus to choose a parameterization -- this is referred to as a choice of \emph{architecture} -- and then find a means of optimizing these parameters $\theta$ with resepect to a chosen \emph{objective}, via a process referred to as \emph{training}.

\subsection{Objective}
\label{sec:objective}

In principle we could determine $\theta$ by maximizing the training set's joint likelihood $\Pi_i p_\theta(\mathbf x^i)$. In practice, however, this would involve evaluating the integral in Equation~\ref{eq:likelihood} for each datapoint individually, which is intractable for even moderately high-dimensional latent spaces. The VAE framework provides an objective function that bounds the maximum likelihood value, and is computationally tractable.

Let a dataset $\mathcal{D}$ be made up of samples $\mathbf{x}^{(i)} = (x_1^{(i)}, \dots, x_N^{(i)}) \in \mathbb{R}^N$, which we will assume to be independent and identically distributed samples from some true underlying distribution $p_{\mathcal{D}}(\mathbf x)$. Absent an analytical model for $p_{\mathcal{D}}(\mathbf x)$, we can instead take it to be a member of an expressive family of functions parametrized by $\bm \theta$: $p_{\mathcal{D}}(\mathbf x) = p_{\bm \theta}(\mathbf{x})$.  This can be done by introducing an unobserved set of latent variables, $\mathbf z = (z_1, \dots, z_d) \in \mathbb{R}^d$, and considering the joint distribution $p(\mathbf x, \mathbf z)$. This joint distribution is specified by: the prior over the latent space, $p(\mathbf z)$, which is assumed to be some simple distribution (typically Gaussian); and the conditional distribution $p(\mathbf x | \mathbf z)$, which is intended to represent most of the complexity in the true underlying distribution $p_{\mathcal{D}}(\mathbf x)$. We model this distribution as a neural network with weights $\theta$: $p_\theta(\mathbf x | \mathbf z)$. The marginal likelihood is then:

\begin{equation}
\label{eq:marginal_likelihood}
    p_\theta(\mathbf{x}) = \int d\mathbf{z}~ p(\mathbf z) p_\theta(\mathbf x | \mathbf z) = 
    \mathbb E_{p(\mathbf z)}\left[ p_\theta(\mathbf x | \mathbf z)\right],
\end{equation}
where we have introduced the notation $\mathbb E_Y[h(y)]$ to indicate the expectation of the function $h(y)$ with respect to the distribution $y \sim Y$. 
In principle, we could determine the conditional model by fixing $\theta$ to a value that maximizes the marginal likelihood. In practice, however, the integral in Equation~\ref{eq:marginal_likelihood} is intractable, due to the dimensionality of the latent space, and in any case would require a per-datapoint optimization process. As a result, the posterior $p_\theta(\mathbf z | \mathbf x) = p_\theta(\mathbf z, \mathbf x) / p_\theta(\mathbf x)$ is also intractable.

We make progress by introducing a second approximation, this time to the posterior: $q_\phi(\mathbf z | \mathbf{x}) \approx p_\theta(\mathbf z | \mathbf{x})$, where $q_\phi(\mathbf z | \mathbf x)$ is often referred to as an \emph{inference} network. For any choice of $q_\phi(\mathbf z | \mathbf x)$, including any choice of its weights $\phi$, we can write the log likelihood of the data as:
\begin{equation}
\label{eq:log_marginal_likelihood}
    \log~p_\theta(\mathbf x) = \mathbb{E}_{q_\phi(\mathbf z | \mathbf x)}\left[\log~p_\theta(\mathbf x) \right] .
\end{equation}
Applying the chain rule of probability: $p_\theta(\mathbf x, \mathbf z) = p_\theta(\mathbf z) p_\theta(\mathbf x | \mathbf z)$, and inserting an identity, this can be split into two terms:
\begin{equation}
\label{eq:log_marginal_likelihood_elbo}
    \log p_\theta(\mathbf x) = \mathbb{L}_{\theta, \phi}(\mathbf x) + 
    \mathbb{D}_{\rm KL}(q_\phi(\mathbf z | \mathbf x) || p_\theta(\mathbf{z} | \mathbf{x})),
\end{equation}
where $\mathbb{L}_{\theta, \phi}$ is referred to as the \emph{evidence lower bound} (ELBO):
\begin{equation}
\label{eq:elbo}
    \mathbb{L}_{\theta, \phi}(\mathbf x) \equiv \mathbb{E}_{q_\phi(\mathbf z | \mathbf x)}\left[\log \left[ \frac{p_\theta(\mathbf x, \mathbf z)}{q_\phi(\mathbf z | \mathbf x)}\right] \right],
\end{equation}
and the second term is the Kullback-Leibler (KL) divergence:
\begin{equation}
\label{eq:kld}
    \mathbb{D}_{\rm KL}(q_\phi(\mathbf z | \mathbf x)||p_\theta(\mathbf z | \mathbf x)) = 
    \mathbb{E}_{q_\phi(\mathbf z | \mathbf x)}\left[ \log \left[ \frac{q_\phi(\mathbf z | \mathbf x)}{p_\theta(\mathbf z | \mathbf x)} \right] \right],
\end{equation}
which is a measure of the `distance' between two distributions, and is always positive. 

From Equation~\ref{eq:log_marginal_likelihood_elbo} we see that the bound $\mathbb{L}_{\theta, \phi}(\mathbf x)$ will become tightest when $\mathbb{D}_{\rm KL}(q_\phi(\mathbf z | \mathbf x)||p_\theta(\mathbf z | \mathbf x))\rightarrow 0$, such that our approximation to the posterior, $q_\phi(\mathbf z | \mathbf x) \approx p_\theta(\mathbf z | \mathbf x)$, becomes exact. However, due to the presence of the $p_\theta(\mathbf z | \mathbf x)$ term, $\mathbb{D}_{\rm KL}(q_\phi(\mathbf z | \mathbf x)||p_\theta(\mathbf z | \mathbf x))$ can not be evaluated directly, and so we are not able to directly optimize the likelihood in Equation~\ref{eq:log_marginal_likelihood_elbo}. Instead, we seek to maximize the evidence lower bound, thereby achieving an `optimum' set of weights $\theta,~\phi$. 

The evidence lower bound and its gradient with respect to $\theta$ can be computed straightforwardly. The gradients with respect to $\phi$ appear more problematic, since the expectation we are calculating is taken over a distribution parametrized by $\phi$. The typical Monte Carlo estimates of this expectation, and its derivatives, are unbiased, but tend to have a high variance, often making the training process unstable. Through a reparametrization presented in \cite{kingma/welling:2013}, it is possible to rewrite this expectation such that the source of randomness is not dependent on $\phi$, and gradients with respect to $\phi$ may be calculated with standard Monte Carlo techniques. We are therefore able to optimize $\mathbb{L}_{\theta, \phi}(\mathbf x)$ by stochastic gradient descent, and approximately optimize the marginal log likelihood. 

\subsection{Architecture}
\label{sec:architecture} 

In this section we describe the architecture of the networks $p_\theta(\mathbf x | \mathbf z)$ and $q_\phi(\mathbf z | \mathbf x)$, and the latent prior $p(\mathbf z)$. We adopt a convolutional architecture for both the encoder and decoder network. 

\subsubsection{Latent Space} We choose to use a $d$-dimensional latent space, with a multivariate normal prior, $\mathbf z \sim \mathcal{N}(0, \mymathbb{1}^{d \times d})$.

\subsubsection{Encoder} The encoder maps input images $\mathbf x \in \mathbb{R}^{256\times 256}$ to latent space distribution parameters, $\mathbf [\bm \mu^d, \bm  \sigma^d] \in \mathbb{R}^{2d}$. It is worth emphasizing the point that, since we are modelling the distribution $p(\mathbf z | \mathbf x)$, the output of the encoder is not a single point in the latent parameter space, but rather a distribution, parametrized by the mean and variance $\mathbf [\bm \mu^d, \bm  \sigma^d]$. The mapping from image to latent space parameters requires both a dimensionality reduction, and a reshaping. We achieve these goals by using a \emph{convolutional neural network}. In the following we will describe the precise network that we implemented, using the language of neural networks. For details on the motivation for these choices, and their technical meaning, we refer to introductory texts on machine learning and convolutional neural networks such as \cite{goodfellow/etal:2016}

The encoder reduces the dimension of the input image by applying a series of strided convolutions with a rectified linear unit activation function, and then flattens the image for input to a final dense layer connected to the output latent space distribution parameters. Each convolution is characterized by a kernel shape with a number of pixels, $k_i$, where $i$ indicates the layer, and a stride length, which we set to 2. The values $k_i$ are set during the hyperaparameter optimization stage described in Section~\ref{sec:hyperparameter_optimization}. We apply a batch normalization with momentum parameter equal to 0.9 after each convolution. This regularizes the weights, and leads to more stable training. A summary of the encoder model is given in Table~\ref{tab:encoder}.

\subsubsection{Decoder} The decoder is essentially the reverse process to the encoder, mapping a latent vector $\mathbf z \in \mathbb R^d$ to an image $\mathbf x \in \mathbb{R}^{256\times256}$. We denote a decoder $g$, with weights $\phi$ as $g_\phi: \mathbf z \rightarrow \mathbf x$. The primary difference to the structure of the encoder is that we use transverse convolutions as opposed to convolutions, in order to increase the size of each dimension. A summary of the decoder model is given in Table~\ref{tab:decoder}.

\begin{table}
    \centering
    \begin{tabular}{c|c|c}
        \hline
        Layer & Layer Output Shape & Hyperparameters \\
        \hline
        \hline
        Input & (256, 256, 1) & \\
        Conv2D & (128, 128, 256) & stride=2\\
        ReLu & (128, 128, 256) & \\
        BatchNorm & (128, 128, 256) & momentum=0.9 \\
        Conv2D & (64, 64, 128) & stride=2 \\
        ReLu & (64, 64, 128) & \\
        BatchNorm & (64, 64, 128) & momentum=0.9 \\
        Conv2D & (32, 32, 64) & stride=2 \\
        ReLu & (32, 32, 64) & \\
        BatchNorm & (32, 32, 64) & momentum=0.9 \\
        Dense & (1024) & \\
        Dense & (512) & \\
        \hline
    \end{tabular}
    \caption{This table shows the structure of the encoder network, $q_\phi(\mathbf z | \mathbf x)$.}
    \label{tab:encoder}
\end{table}

\begin{table}
    \centering
    \begin{tabular}{c|c|c}
         \hline
        Layer & Layer Output Shape & Hyperparameters \\
        \hline
        \hline
        Input & (256, 1) & \\
        Dense & (8192) & \\
        Reshape & (16, 16, 32) & \\
        BatchNorm & (16, 16, 32) & momentum=0.9 \\
        TransposeConv2D & (32, 32, 128) & stride=2\\
        ReLu & (32, 32, 128) & \\
        BatchNorm & (32, 32, 128) & momentum=0.9 \\
        TransposeConv2D & (64, 64, 64) & stride=2\\
        ReLu & (64, 64, 64) & \\
        BatchNorm & (64, 64, 64) & momentum=0.9 \\
        TransposeConv2D & (128, 128, 32) & stride=2\\
        ReLu & (128, 128, 32) & \\
        BatchNorm & (128, 128, 32) & momentum=0.9 \\
        TransposeConv2D & (256, 256, 16) & stride=2\\
        ReLu & (256, 256, 16) & \\
        BatchNorm & (256, 256, 16) & momentum=0.9 \\
        TransposeConv2D & (256, 256, 1) & stride=1\\
        \hline
    \end{tabular}
    \caption{This table shows the structure of the decoder network, $p_\theta(\mathbf x | \mathbf{z})$.}
    \label{tab:decoder}
\end{table}

\subsection{Training}
\label{sec:training} 

In this section we detail the process by which we optimize the weights of the VAE model described in Section~\ref{sec:architecture} with respect to the ELBO objective introduced in Section~\ref{sec:objective}. 
The training process requires us to specify the training dataset, $\mathcal{D}$, the training \emph{strategy} by which we make updates to the weights $\theta,~\phi$, and the process of hyperparameter optimization by which we make concrete selections of meta parameters of the model (such as kernel shapes and training parameters).

\subsubsection{Data}
\label{sec:data} 

Machine learning techniques are notoriously data-hungry, and will perform best for larger datasets. Standard computer vision datasets on which algorithms are tested (e.g. ImageNet \citep{imagenet}) contain tens of thousands, sometimes millions, of images. However, we have only one sky from which to obtain observations of Galactic dust. As such, we are forced to partition the sky into patches, which we treat as separate images in the training process. In order to obtain $\sim 1000$'s of images, the natural linear scale of an individual patch is $\sim 10^\circ$. Such a small patch size has the advantage that we are then justified in projecting the cutouts onto the flat sky, and applying standard machine learning techniques to the resulting two-dimensional images, sidestepping the issue of defining neural networks that operate on spherical images (for such implementations see \cite{perraudin/etal:2019,krachmalnicoff/tomasi:2019}). 

We use the {\emph Planck} GNILC-separated thermal dust intensity map at 545 GHz \footnote{{\url{http://pla.esac.esa.int/pla/aio/product-action?MAP.MAP_ID=COM_CompMap_Dust-GNILC-F545_2048_R2.00.fits}}}, which we download from the Planck Legacy Archive. In order to extract cutout images from this map we follow a similar procedure to \cite{aylor/etal:2019}. We mask the Galactic plane by excluding all regions at latitudes below $15^\circ$. Then we lay down a set of centroids $(l_{i+1}, b_{i+1}) = (l_i + s, b_i + s / \cos(l_i))$, where $s$ is a step size parameter, and $s/\cos(l_i)$ is a step between longitudes for a given latitude, which ensures the same angular separation in the latitudinal direction. Each centroid is then rotated to the equator, and an $8^\circ \times 8^\circ$ square region around the centroid is projected onto a cartesian grid with 256 pixels along each size. For $s=4^\circ$, this results in a dataset, $\mathcal D$, of 2254 maps. We then shuffle and split $\mathcal D$ into three groups: a 70\% training set, $\mathbf x^{\rm train}$, a 15\% validation set, $\mathbf x^{\rm val}$, and a 15\% test set, $\mathbf x^{\rm test}$. 

In order to artificially increase the diversity of images in our limited sample we employ two standard data augmentation techniques. During the data preprocessing stage of training, we randomly flip each image along the horizontal and vertical directions, and rotate each image by an integer multiple of $90^{\circ}$. These transformations are not invariant under convolution; however, these would constitute perfectly realistic foreground images.

\subsubsection{Strategy}
\label{sec:schedule} 

Here we discuss the training strategy used to learn the weights $\theta, \phi$.

As discussed in Section~\ref{sec:variational_autoencoders}, to train a VAE we maximize the lower bound on the log likelihood of the data given in Equation~\ref{eq:elbo} with respect to the weights $\theta, \phi$. In practice, at each step we compute a Monte Carlo estimate of this quantity:

\begin{equation}
    \mathbb{E}_{q_\phi(\mathbf z | \mathbf x)}\left[\frac{p_\theta(\mathbf x , \mathbf z)}{q_\phi(\mathbf z | \mathbf x)} \right] \approx \log p_\theta(\mathbf x | \mathbf z) + \log p(\mathbf z) - \log q_\phi(\mathbf z | \mathbf x)
\end{equation}

where $\mathbf x$ on the RHS is now a minibatch of the data, the size of which is a hyperparameter of the training process. The analysis we present in Section~\ref{sec:hyperparameter_optimization} shows that a batch size of 8 is preferred. For each batch we then calculate the gradients of this quantity with respect to the weights $\theta, \phi$ and backpropagate the errors through the network, adjusting $\theta, \phi$ in accordance with the learning schedule. For this schedule we used the Adam optimizer with hyperparameters determined through the optimization process described in Section~\ref{sec:hyperparameter_optimization}.

The training was performed by passing over the entire dataset 100 times, and in each pass splitting the data into batches of 8 images. To guard against overfitting we evaluated $\mathbb{L}_{\theta, \phi}(\mathbf x^{\rm train})$ and $\mathbb{L}_{\theta, \phi}(\mathbf x^{\rm val})$ every five epochs and checked for divergence between these quantities at late epochs. If the network had begun to overfit on the training data, its predictions for the validation set would deteriorate, which would be reflected in a worsening $\mathbb{L}_{\theta, \phi}(\mathbf x^{\rm val})$. We found that the $\mathbb{L}_{\theta, \phi}(\mathbf x^{\rm train})$ plateaued after 50 epochs, and saw no divergence between $\mathbb{L}_{\theta, \phi}(\mathbf x^{\rm train})$ and $\mathbb{L}_{\theta, \phi}(\mathbf x^{\rm val})$ after training for an additional 50 epochs. 

 Models were built using the Tensorflow software package \citep{tensorflow}, and trained using a Tesla V100 GPU on the Cori supercomputer at NERSC.

\subsubsection{Hyperparameter Optimization}
\label{sec:hyperparameter_optimization} 

In this section we provide motivation for our selection of the model hyperparameters. It is not possible to optimize model hyperparameters such as batch size, or model architecture, using the same stochastic gradient descent technique that is used to optimize model weights and biases. Instead, a limited number of hyperparameter combinations can be trained, and the corresponding model that achieves the best loss after a certain amount of training time, or certain number of epochs, is used. The space of hyperparameters is high-dimensional, and so can not be uniformly densely sampled due to computational cost. Instead, we employed a Bayesian optimization approach in which a few random combinations of hyperparameters are chosen, and trained for 20 epochs each. From this set of hyperparameters, a Gaussian process (GP) model of the loss as a function of hyperparameters is built. From this GP model, new trial candidates are selected, and trained, with the resulting loss then being incorporated into the GP weights. We allowed this process to continue for 100 different trials, and used the hyperparameters that achieved the lowest loss after twenty epochs of training. 

\section{Results}
\label{sec:results} 

\subsection{Reconstructions}
\label{sec:recall} 

In this section we present reconstructions of test set images, and compare their pixel value distribution and power spectra. 

For a given image, $\mathbf x_{\rm test}$, we can sample the posterior as $\mathbf z_{\rm test}^{(i)} \sim q_\phi(\mathbf z | \mathbf x)$, and push these through the decoder to get a reconstructed image $\mathbf x^{(i)}_{\rm test} = \mathbf g_\theta(\mathbf z^{(i)}_{\rm test})$. To summarize the distribution of reconstructed images, we draw $L$ samples and calculate their average:
\begin{equation}
\label{eq:reconstruction}
    \tilde{\mathbf x} \approx \frac{1}{L} \sum_{l=1}^{L} \mathbf g_\theta(\mathbf z_{\rm test}^{(l)}).
\end{equation}
For the remainder of this section, a `reconstruction' refers to the calculation of Equation~\ref{eq:reconstruction} with $L=100$. For a given reconstruction, we can straightforwardly calculate two statistics: i) the histogram of its pixel values and ii) the power spectrum. We calculate the histogram of pixel values in 20 bins from -3 to 5, and normalize the count such that the area under the histogram is equal to unity. To calculate the power spectrum we apply a cosine apodization with a characteristic scale of one degree to the image, such that it smoothly tapers to zero at the edge of the map. We then calculate the mode coupling matrix for this mask, and calculate the uncoupled power spectrum using the \href{https://github.com/LSSTDESC/NaMaster}{\tt NaMaster} code \citep{alonso/etal:2019}. For reasons that will become clear later we are primarily interested in comparing ranges of multipoles in the signal-dominated regime, well within the resolution limit of the original maps, and so we do not make any efforts to noise debias or account for the beam present in the original maps.

\begin{figure*}
    \centering
    \includegraphics{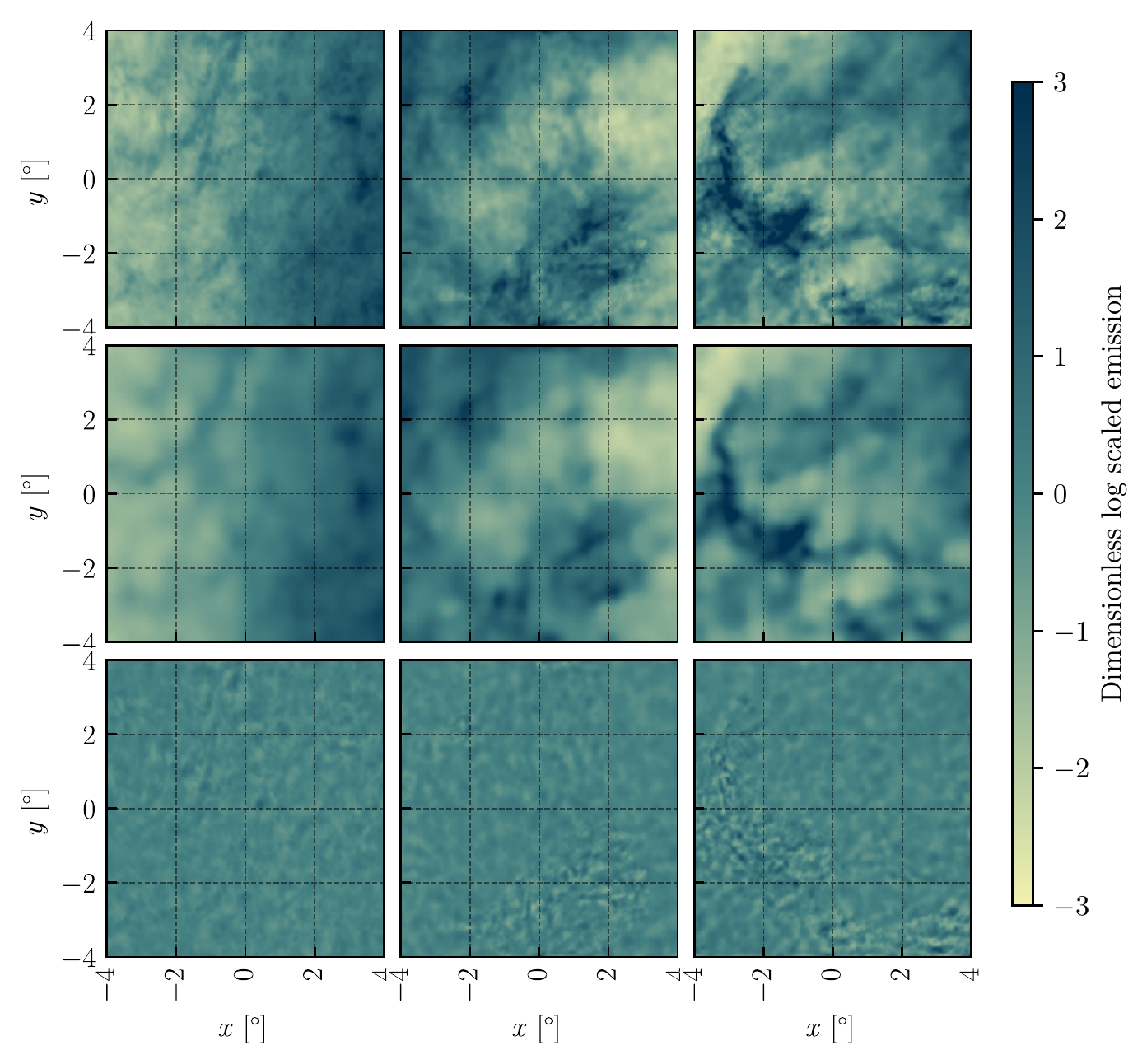}
    \caption{This figure shows the reconstruction of three randomly-selected images from the test set, not used during the training or validation of the network. The top row are the original images, the second row are the reconstructions. and the third row are the residuals of the reconstructions. The reconstructions clearly lose small-scale details, but but manage to recover the large scale variations well.}
    \label{fig:recon_w_res}
\end{figure*}

First, we present the reconstructions of three randomly-selected test set images, and show the resulting maps, along with the residuals, in Figure~\ref{fig:recon_w_res}. We can see that the network does very well in reconstructing the large-scale features in these test-set maps, and the visual quality is sufficient to appear `real', if lower-resolution. Features are well recovered up to $\sim$degree scales, with features below that scale being smoothed out by the calculation of the expectation in Equation~\ref{eq:reconstruction}. The residuals shown in the bottom row of Figure~\ref{fig:recon_w_res} are well behaved and do not show any strong biases correlated with features in the map. 

\begin{figure*}
    \centering
    \includegraphics{ 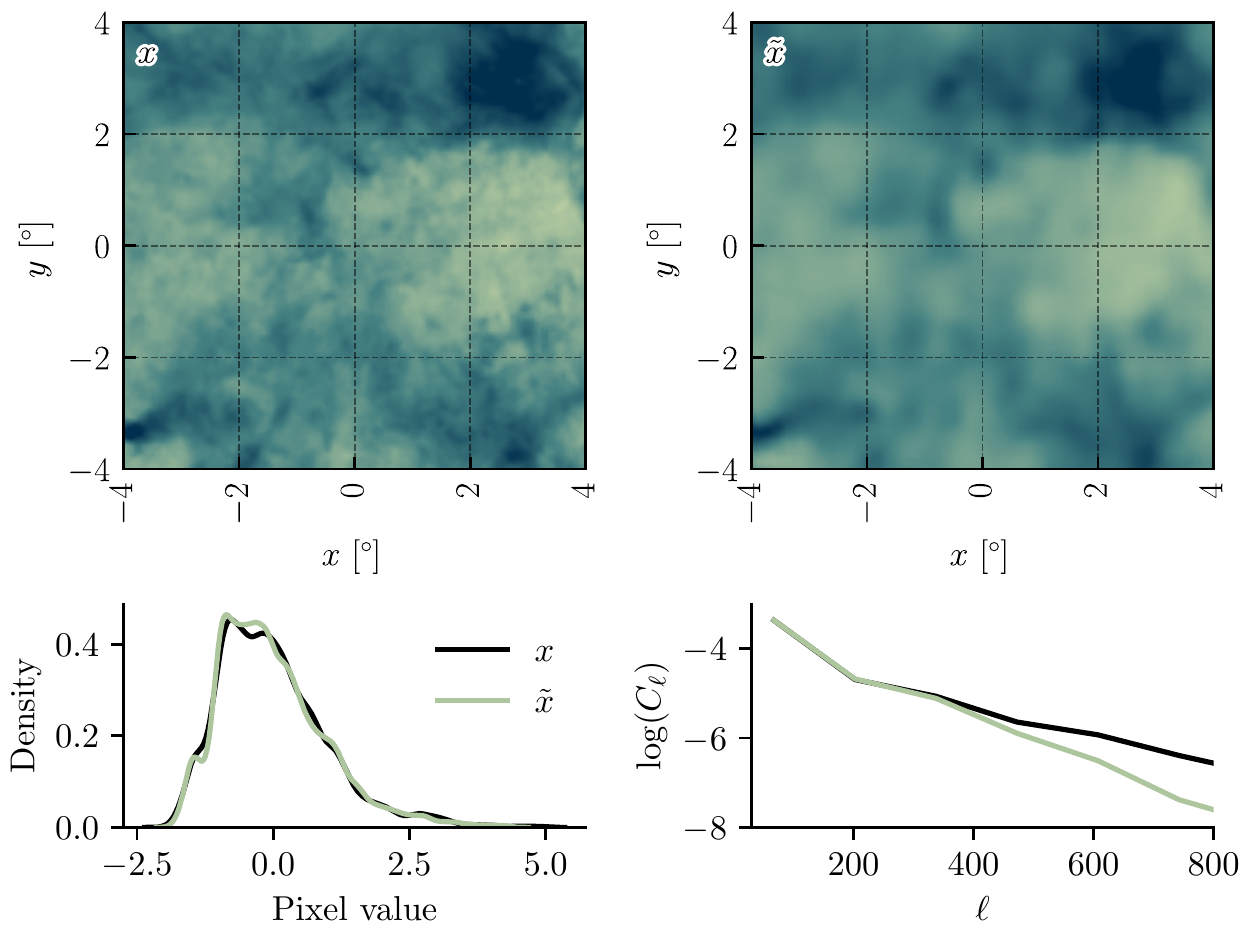}
    \caption{\emph{Top left}: a randomly-selected test set image, $\mathbf x$. \emph{Top right}: the reconstruction of the test set image, $\tilde{\mathbf x}$, as computed using Equation~\ref{eq:reconstruction}. \emph{Bottom left}: kernel density estimate of the distribution of pixel values of the original image, and its reconstruction. \emph{Bottom right}: the log power spectra of the test set image and its reconstruction. Note that since the test set images are standardized, these quantities are unitless.}
    \label{fig:panel_plot}
\end{figure*}

In Figure~\ref{fig:panel_plot} we take a single randomly-selected test set image, and show its reconstruction, the pixel value histograms of each image, and their power spectra. As was the case for the three examples shown in Figure~\ref{fig:recon_w_res}, there is excellent visual agreement between the original image and its reconstruction. This is enforced by the excellent agreement between the distribution of pixel values in the two images, shown in the bottom left panel of Figure~\ref{fig:panel_plot}. The reconstructed power spectrum in the bottom right panel of Figure~\ref{fig:panel_plot} also shows excellent agreement up to $\ell \sim 400$, and suppression of power in the reconstructed image going to smaller scales, consistent with the visual blurriness of the reconstructed image.

\begin{figure}
    \centering
    \includegraphics{ 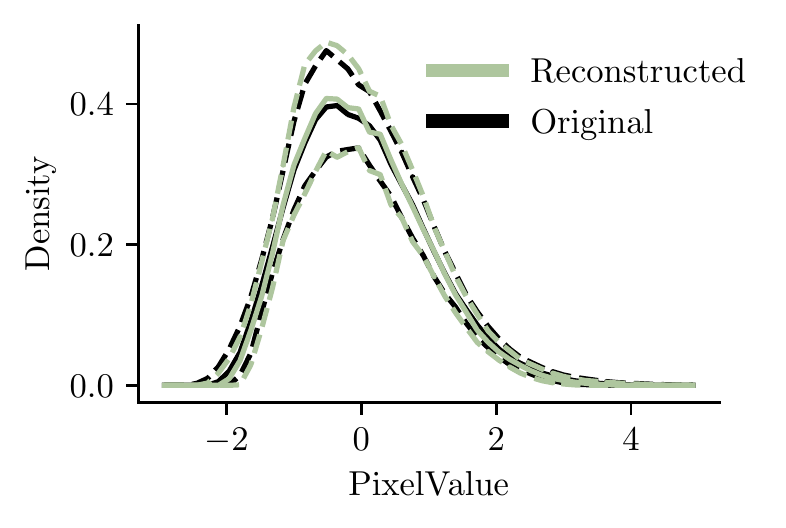}
    \caption{In this figure we compare the pixel value distributions of the 339 test set images (black), and their reconstructions (green). We calculate quantiles across the test set, and plot the $25^{\rm th}$ and $75^{\rm th}$ quartiles (the dashed lines), and the median as functions of pixel value (the solid lines).}
    \label{fig:histogram_distribution}
\end{figure}

\begin{figure}
    \centering
    \includegraphics{ 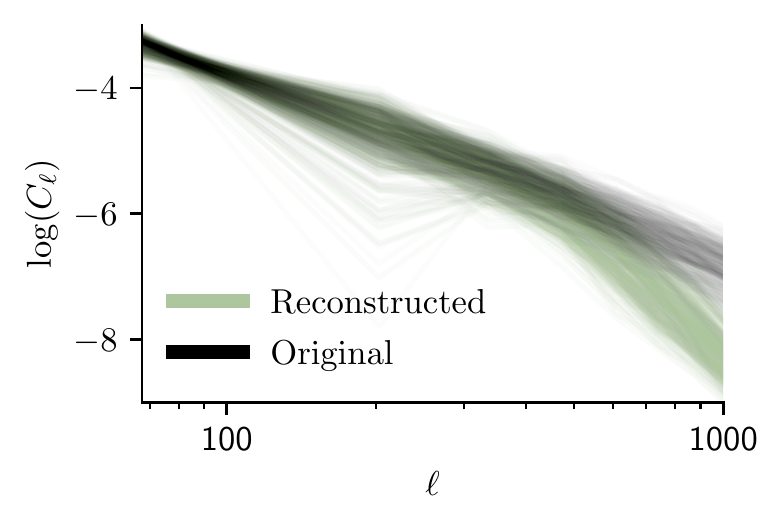}
    \caption{In this figure we compare the power spectra of the 339 test set images (black) and their reconstructions (green). Each power spectrum is plotted as an individual line.}
    \label{fig:powerspectrum_distribution}
\end{figure}

In order to compare reconstructions for the whole test set, we now calculate the pixel value distribution and power spectrum for each of the 339 images in the test set and their reconstructions. In order to represent the distribution of pixel value histograms across this test set, we calculate the quartiles and median in each bin, across the test set. In Figure~\ref{fig:histogram_distribution} we plot the $25^{\rm th}$ percentile, median, and $75^{\rm th}$ percentile as a function of bin center, for both the original test set images, and their reconstructions. There is excellent agreement between the two sets of images, with no evidence of any aggregate bias in the reconstructions.
In Figure~\ref{fig:powerspectrum_distribution} we compare the power spectra of all test set images and their reconstructions. Figure~\ref{fig:powerspectrum_distribution} shows that the same behavior as was seen in Figure~\ref{fig:panel_plot} is displayed for the entire test set. Spectra are generally well recovered for $\ell < 400$, with power being increasingly suppressed for $\ell > 400$, relative to the real image power spectra.

Here, we are encountering a known issue with VAEs: reconstructed images are often blurry \citep{kingma/dhariwal:2018, kingma/etal:2016, kingma/welling:2019}. The blurriness can be understood by considering the objective function in Equation~\ref{eq:elbo}, and inspecting the term $\mathbb E_{q_\phi(\mathbf z | \mathbf x)}\left[p_\theta(\mathbf x, \mathbf z) \right]$. Since this expectation is taken with respect to the distribution $q_\phi(\mathbf z | \mathbf x)$, it will strongly penalize points $(\mathbf x, \mathbf z)$ that are likely under $q_\phi$, but unlikely under $p_\theta$. On the other hand, points that are likely under $p_\theta$, but are not present in the empirical data distribution, will suffer a much smaller penalty. The result is that, if the model is not sufficiently flexible to fit the data distribution exactly, it will compensate by widening the support of $p_\theta(\mathbf x, \mathbf z)$ beyond what is present in the data distribution, inflating the variance of $p_\theta(\mathbf x | \mathbf z)$. Since we have assumed a Gaussian distribution for the decoder model that is independent from pixel to pixel, and given that the signal in the training images is red-tilted (as is the case for most natural images containing extended recognizable structures), the increased variance leads to a degradation of small-scale features through the averaging process of Equation~\ref{eq:reconstruction} \citep{zhao/etal:2017}. 
A corollary of the extended support of $p_\theta(\mathbf x, \mathbf z)$ is that sampling the prior in order to generate novel images will not necessarily produce realistic samples \citep{kingma/welling:2019}.  

One way in which the flexibility of VAEs may be enhanced is through the use of \emph{normalizing flows} \citep{rezende/etal:2015}. As the name suggests, the idea here is to start with a simple distribution, such as a multivariate normal, and `stack' layers of invertible transformations, such that the output may be significantly more complex. There are certain requirements placed on these transformations such that they remain computationally efficient, for example they must have tractable Jacobians \citep{rezende/etal:2015}.  
Expanding the VAE model presented here by introducing normalizing flows could be expected to improve both the reconstruction quality, and the quality of novel samples, and is the subject of current work.

\subsection{Interpolation in the latent space}
\label{sec:latent_interpolation} 
As a means of investigating the structure of the encoding that has been learned, we study the `interpolation' between real images, $\mathbf x_1$ and $\mathbf x_2$, by performing the interpolation between their latent encodings, $\mathbf z_1$ and , $\mathbf z_2$. From the smooth nature of the changes in the resulting continuum of maps we will see that smooth variations in the latent space result in smooth variations in the map space. This study also demonstrates the ability of the VAE approach to generate novel foreground images by restricting to a region of the latent space close to the encodings of real maps, therefore avoiding the spurious regions of $(\mathbf x, \mathbf z)$ that could be obtained by sampling from an ill-fitted prior, as discussed at the end of Section~\ref{sec:recall}.

The probability mass in high-dimensional distributions tends to concentrate in a shell relatively far from the modal probability density. Therefore, traversing the latent space in a straight line (in the Euclidean sense), does not necessarily pass through areas of high probability mass. In order to keep the interpolated points within areas of high probability mass, we interpolate from $\mathbf z_1$ to $\mathbf z_2$ using spherical trajectories that traverse great circles in the latent space, as the distance from the origin smoothly changes from $|\mathbf{z_1}|$ to $|\mathbf{z_2}|$. Specifically, we follow this continuous trajectory parametrized by some factor $\lambda$: 
\begin{equation}
\label{eq:interpolation}
    \mathbf{z}_{1,2}(\lambda) = \frac{\sin((1- \lambda) \theta)}{\sin \theta} \mathbf z_1 + \frac{\sin(\lambda \theta)}{\sin \theta} \mathbf z_2,
\end{equation}
where $\cos(\theta) = \hat{\mathbf z}_1 \cdot \hat{\mathbf z}_2$. We then take $N$ points along this line corresponding to $\lambda = [1/(N+1), 2/(N+1), \dots, N/(N+1)]$, and decode to obtain the corresponding map $\mathbf x_{1,2}(\lambda) = g_\phi(\mathbf z_{1,2}(\lambda))$.

\begin{figure*}
    \centering
    \includegraphics{ 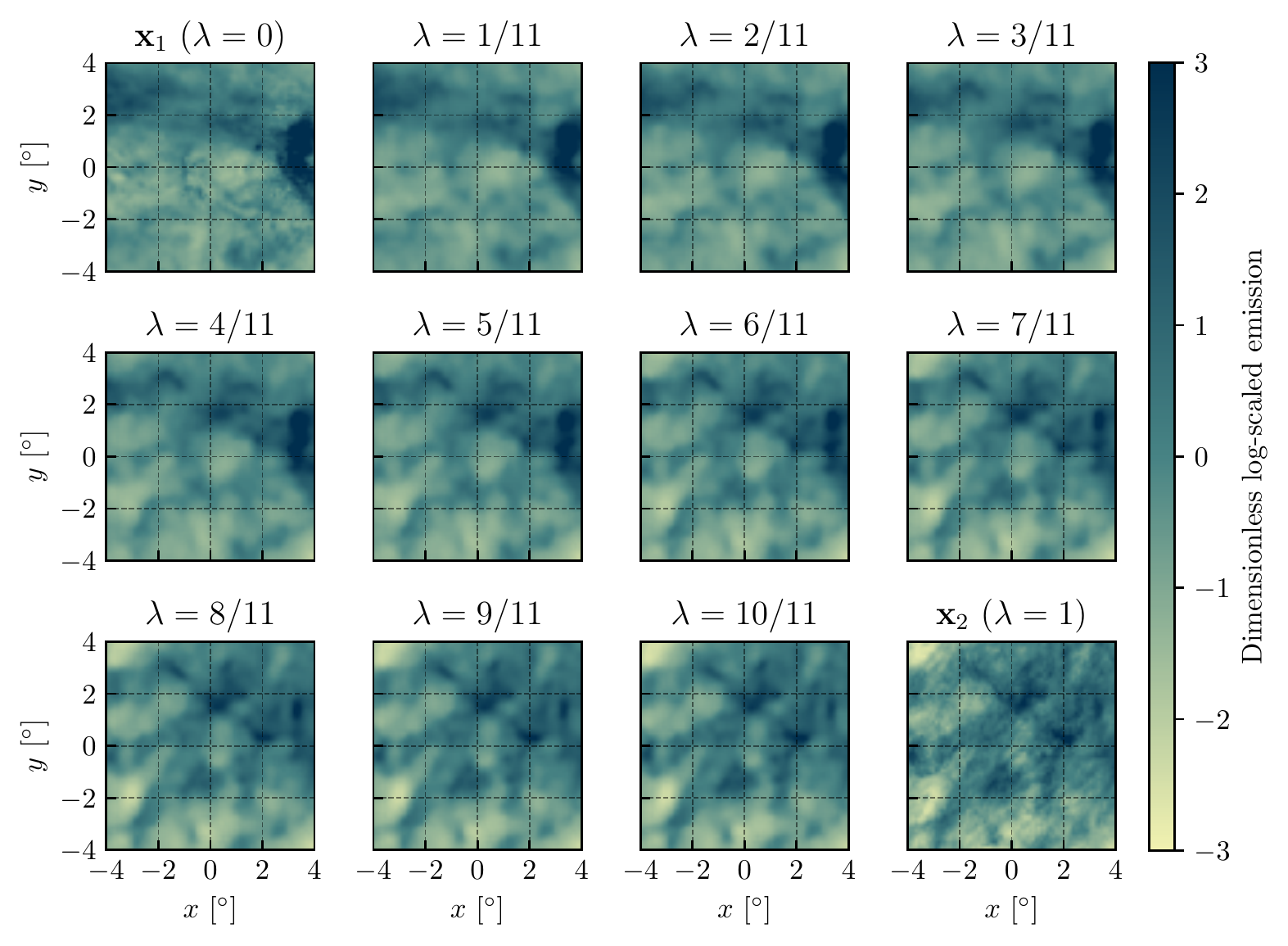}
    \caption{This figure presents synthetic images generated by interpolating between real images, $\mathbf x_1$ and $\mathbf x_2$, shown in the top left and bottom right panels respectively. The interpolation is carried out in the latent space using Equation~\ref{eq:interpolation}, and is parametrized by a continuous variable $\lambda$. The intermediate panels show the interpolation evaluated at $N=10$ points along the trajectory.}
    \label{fig:interpolation}
\end{figure*}

Figure~\ref{fig:interpolation} shows the smooth transition in image space between the two real images (the top left panel and the bottom right panel) randomly selected from the test set, calculated using the interpolation described above. Features, such as the strong filamentary structures in the center of the image, transition smoothly in and out of the image, demonstrating that small perturbations in the latent space result in small perturbations in decoded images.

\subsection{Data Imputation}
\label{sec:imputation} 
In this section we consider a possible application of our trained model to the reconstruction of corrupted data. During the analysis of CMB data there are many possible reasons that data may be incomplete, from masking of point sources, to corruption by uncontrolled systematics. The task of \emph{inpainting} these regions is simple when the missing emission is well described by Gaussian statistics, as is the case for the CMB \citep{bucher/louis:2012}. The lack of a similarly simple approach for the non-Gaussian foreground signal means that previous efforts have relied on empirically-validated, simple, algorithms, such as diffusive filling \citep{bucher/etal:2016}. Future surveys will have ever-lower noise floors, and so will be increasingly contaminated by point-sources, even in polarization. The aggressive masking required in this regime could lead to the failure of simple foreground inpainting techniques \citep{puglisi/bai:2020}. The statistical foreground model presented here allows us to take a Bayesian approach to foreground inpainting, in which we may compute a posterior distribution for the missing data, conditioned on the observed data \citep{bohm/seljak:2019}. This has the advantage of conserving the foregrounds' statistical properties, whilst also taking into account all of the contextual information in the image, unlike methods such as diffusive inpainting. In the rest of this section we will present a toy model for corrupted data, and show that we are able to perform inpainting by optimizing the posterior distribution in the latent space.

Representing the contamination as a linear operator $\mathsf A$, we can write down a model for the observed data $\mathbf d$: $\mathbf d = \mathsf A \mathbf x + \mathbf n$, where $\mathbf n$ is a possible noise term. The posterior distribution of $\mathbf z$ is given by Bayes' theorem:
\begin{equation}
    \log p(\mathbf z | \mathbf d) = \log p(\mathbf z) + \log p_\theta(\mathbf d | \mathbf z) - \log p(\mathbf d).
\end{equation}
For a given statistical model of the noise, we have a complete description of the term $\log p(\mathbf d | \mathbf z)$, and we can work with the posterior distribution in the latent space. 

As a concrete example we will consider the case of a binary $N\times N$ masking operator, $\mathsf A$, with elements equal to one (zero) where pixels are (un)observed. To form simulated `corrupted' images, we take random images from the test dataset, apply $\mathsf A$, and add white Gaussian noise $\mathbf n$, characterized by a pixel standard deviation $\sigma$: $\mathbf d_{\rm test} = \mathsf A \mathbf x_{\rm test} + \mathbf n$. The posterior distribution in the latent space is then:
\begin{equation}
\label{eq:imputation_posterior}
  - 2 \log p(\mathbf z | \mathbf d_{\rm test}) \propto \mathbf z^T \mathbf z + \frac{\bm \mu_\theta(\mathbf z)^T \bm \mu_\theta(\mathbf z)}{\sigma^2},
\end{equation}
where we have written the residual vector as $\bm \mu_\theta(\mathbf z) = \mathsf A \mathbf g_\theta(\mathbf z) - \mathbf d_{\rm test}$.

Fully sampling Equation~\ref{eq:imputation_posterior} can be computationally expensive due to the dimensionality of $\mathbf z$, and is made more challenging by the possibility of $\log p(\mathbf z| \mathbf d_{\rm test})$ being multi-modal. For these reasons, applying standard Markov Chain Monte Carlo techniques can often fail to fully explore the posterior \citep{bohm/seljak:2019}, and we leave a sampling approach for future work, here taking only a single representative sample by maximizing $\hat{\mathbf{z}}_{\rm test} = \argmax_{\mathbf z} \log p(\mathbf z | \mathbf d_{\rm test})$. 

\begin{figure*}[p]
    \centering
    \includegraphics{ 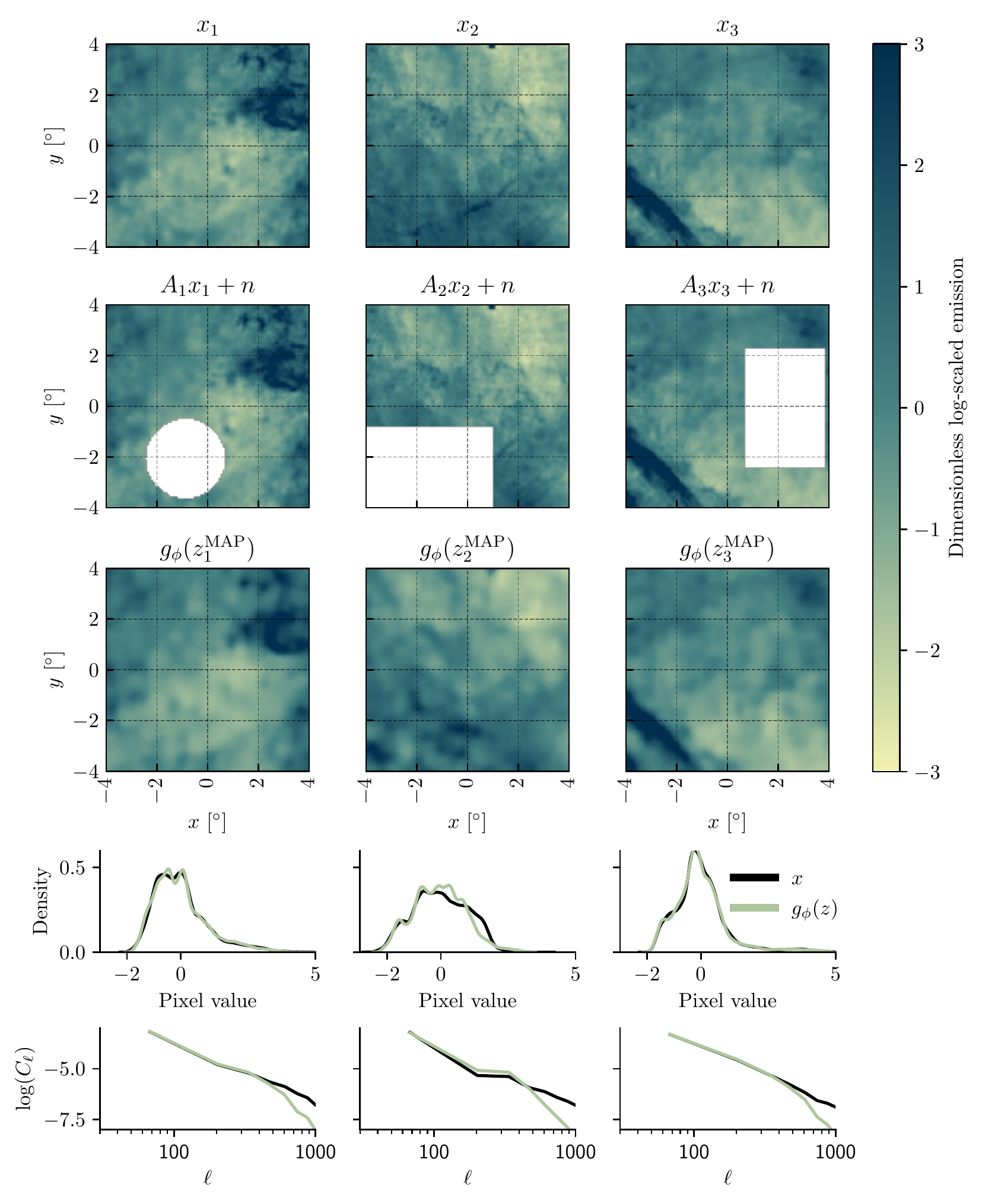}
    \caption{This figure shows three randomly-selected test set images, $\mathbf x_{1, 2, 3}$ in the top row. As described in Section~\ref{sec:imputation}, these images are corrupted with a binary mask $\mathsf A_{1, 2, 3}$ and white noise. The corrupted images are shown in the second row. The third row shows the reconstructed images obtained by maximizing the latent space posterior in Equation~\ref{eq:imputation_posterior} for each of the three corrupted images, and decoding the resulting points in the latent space. The fourth and fifth rows show the pixel value histograms and power spectra of the original and reconstructed maps.}
    \label{fig:data_imputation} 
\end{figure*}

In the following we will take $\mathsf A$ to be a masking operator that applies a binary mask to a map. However, as long as a forward model for the corruption operation can be written down (e.g. a Gaussian convolution), the same technique could be applied. We take three randomly selected test set images, $\mathbf x_1, \mathbf x_2, \mathbf x_3$, and apply three different binary masks, $\mathsf A_1, \mathsf A_2, \mathsf A_3$. To each corrupted image, we add a white noise realization with a pixel standard deviation of 0.2. For each corrupted, noisy image, we then maximize the posterior in Equation~\ref{eq:imputation_posterior} to find $\mathbf z_i^{\rm MAP}$ using the LBFGS algorithm. In Figure~\ref{fig:data_imputation} we show the randomly selected test set images in the first row, the corrupted images in the second row, and the reconstructed map $g(\mathbf z_i^{\rm MAP})$ in the third row. We also calculate the pixel value histograms and power spectra of the input and reconstructed maps and show these in the bottom two rows of Figure~\ref{fig:data_imputation}.

One can see from Figure~\ref{fig:data_imputation} that all the images are well reconstructed, and there is no visible effect of the masking remaining in the reconstructions. Comparing the regions in the first and third rows corresponding to the masked areas, we see that the network does not reproduce the exact features in the masked region, for any of the $\mathbf x_i$, as expected. However, the network does reconstruct plausible inpaintings, with the correct statistics, given the context in the rest of the image. For example, the reconstruction $g_\phi(\mathbf z_2^{\rm MAP})$ does not replicate the true high-intensity filamentary structure in the input image, $\mathbf x_2$, which would be impossible. However, it does recognize from the context that intensity is increasing towards the masked area in the bottom left of the image, and populates that area with high-variance, high-intensity features. Correspondingly, such high-intensity features are not seen in the reconstructed regions of $g_\phi(\mathbf z_{1, 3}^{\rm MAP})$, which correspond to relatively low-emission regions. The pixel value histograms and power spectra in the last two rows of Figure~\ref{fig:data_imputation} show similar behavior. We see good agreement between the original and reconstructed histograms and powerspectra for both the $\mathbf x_1$ and $\mathbf x_3$ maps, up to the suppression at $\ell > 400$ common to all reconstructions. On the other hand, we see a disagreement between the original and reconstructed statistics of $\mathbf x_2$, due to the higher variance associated with the filled-in region.

These results show that the network has learned generalizable information about foreground behavior, and is able to inpaint novel foreground emission with correct statistical properties, based on the context of an image. The forward model used in this inpainting process can be easily extended to maps with multiple masks and different types of filtering and noise found in real data.

\section{Discussion and Conclusions}
\label{sec:conclusions} 

In this paper we have presented a new application of VAEs to images of Galactic thermal dust emission. Using a training set extracted from Planck observations of thermal dust emission, this technique allowed us to learn a transformation from a space of uncorrelated latent variables with a multivariate normal prior, to the space of possible dust maps.
 
The training process was validated by computing and comparing summary statistics, including the distribution of pixel values, and power spectra of reconstructed maps, on a test set withheld during the training process. The applicability of the trained model was also demonstrated by reconstructing data corrupted by noise and masking. This was the first use of a trained generative dust model to perform Bayesian inference, and demonstrates the applicability of this approach in the simulation of foreground images, and the Bayesian modeling of polarized CMB data. 

The usefulness of this model is currently limited by the flexibility of the posterior, and its ability to fit the true underlying posterior. As was discussed in Section~\ref{sec:recall}, this has two main consequences: i) a na\"ive sampling of the prior is not guaranteed to produce realistic samples, ii) reconstructed images are blurry, limiting accuracy to degree scales. Both of these issues may be tackled by increasing the expressiveness of the model \citep{kingma/welling:2019}, which we plan to do by introducing a normalizing flow to link the prior and latent space \citep{kingma/etal:2016}.

As discussed in the Section~\ref{sec:introduction}, our main goal is to model polarized dust emission. We attempted a similar analysis to that presented here by repeating the training procedure on a network that accepted an additional `channel' as input, representing a tuple of Stokes $Q$ and $U$ parameters, rather than only Stokes $I$, and using the Planck 353 GHz polarization observations to form a training set. We found that the network was not able to learn any meaningful information from this setup, consistent with what similar analyses have found \citep{petroff/etal:2020}. In order to extend our analysis to polarization, we are therefore exploring the use of MHD simulations \citep{kim/etal:2019} as a training set. \cite{kim/etal:2019} have demonstrated that simulations of a multiphase, turbulent, magnetized ISM produce synthetic observations of the ISM with statistics (such as the ratio of $E$ power to $B$ power, and the tilt of the $EE$ and $BB$ power spectra) matching those of real skies. Our initial results have shown that this is a promising alternative to the use of real data in training generative networks.

\section*{Acknowledgements}
We would like to acknowledge useful conversations with Ethan Anderes and Kevin Aylor in the preparation of this work. This work was supported by an XSEDE start up allocation, PHY180022. This work was supported in part by the National Science Foundation via awards OPP-1852617 and AST-1836010. 
We also acknowledge the use of the Perlmutter preparedness GPU allocation on the Cori super computer at NERSC.

\section*{Data Availability}

The data used in this study is available on the Planck Legacy Archive at the URL: \url{http://pla.esac.esa.int/pla/aio/product-action?MAP.MAP_ID=COM_CompMap_Dust-GNILC-F545_2048_R2.00.fits}

%%%%%%%%%%%%%%%%%%%% REFERENCES %%%%%%%%%%%%%%%%%%

% The best way to enter references is to use BibTeX:

\bibliographystyle{mnras}
\bibliography{library} % if your bibtex file is called example.bib

% Alternatively you could enter them by hand, like this:
% This method is tedious and prone to error if you have lots of references
%\begin{thebibliography}{99}
%\bibitem[\protect\citeauthoryear{Author}{2012}]{Author2012}
%Author A.~N., 2013, Journal of Improbable Astronomy, 1, 1
%\bibitem[\protect\citeauthoryear{Others}{2013}]{Others2013}
%Others S., 2012, Journal of Interesting Stuff, 17, 198
%\end{thebibliography}

%%%%%%%%%%%%%%%%%%%%%%%%%%%%%%%%%%%%%%%%%%%%%%%%%%

%%%%%%%%%%%%%%%%% APPENDICES %%%%%%%%%%%%%%%%%%%%%

%%%%%%%%%%%%%%%%%%%%%%%%%%%%%%%%%%%%%%%%%%%%%%%%%%

% Don't change these lines
\bsp	% typesetting comment
\label{lastpage}
\end{document}